*Research Article*

# Computer-Aided Decision Support for Melanoma Detection Applied on Melanocytic and Nonmelanocytic Skin Lesions: A Comparison of Two Systems Based on Automatic Analysis of Dermoscopic Images

**Kajsa Møllersen,[1] Herbert Kirchesch,[2] Maciel Zortea,[3] Thomas R. Schopf,[1] Kristian Hindberg,[3] and Fred Godtliebsen[3]**

[1]*Norwegian Centre for Integrated Care and Telemedicine, University Hospital of North Norway, 9038 Tromsø, Norway*
[2]*Private Office, Venloer Straße 107, 50259 Pulheim, Germany*
[3]*Department of Mathematics and Statistics, UiT The Arctic University of Norway, 9037 Tromsø, Norway*

Correspondence should be addressed to Kajsa Møllersen; kajsa.mollersen@telemed.no

Commercially available clinical decision support systems (CDSSs) for skin cancer have been designed for the detection of melanoma only. Correct use of the systems requires expert knowledge, hampering their utility for nonexperts. Furthermore, there are no systems to detect other common skin cancer types, that is, nonmelanoma skin cancer (NMSC). As early diagnosis of skin cancer is essential, there is a need for a CDSS that is applicable to all types of skin lesions and is suitable for nonexperts. Nevus Doctor (ND) is a CDSS being developed by the authors. We here investigate ND's ability to detect both melanoma and NMSC and the opportunities for improvement. An independent test set of dermoscopic images of 870 skin lesions, including 44 melanomas and 101 NMSCs, were analysed by ND. Its sensitivity to melanoma and NMSC was compared to that of Mole Expert (ME), a commercially available CDSS, using the same set of lesions. ND and ME had similar sensitivity to melanoma. For ND at 95% melanoma sensitivity, the NMSC sensitivity was 100%, and the specificity was 12%. The melanomas misclassified by ND at 95% sensitivity were correctly classified by ME, and vice versa. ND is able to detect NMSC without sacrificing melanoma sensitivity.

## 1. Introduction

Melanoma is the deadliest of all skin cancers. When it is detected early, the treatment is excision of the tumour, and the survival rate is high. Recent developments in melanoma treatment are promising [1, 2], but the survival rate for patients with metastasised melanoma is still poor [3, 4]. Early diagnosis is crucial, but challenging, since early stage melanomas resemble benign skin lesions. Other types of skin cancer, like basal cell carcinoma (BCC) and squamous cell carcinoma (SCC), have high incidence rates but low mortality rates [4]. Early detection is beneficiary for the patient to get early treatment and avoid further damaging of the surrounding skin.

The dermoscope (dermatoscope, epiluminescence microscope) reveals structures not visible to the naked eye and has proven to raise diagnostic accuracy of melanoma when used by properly trained personnel [5]. Dermoscopy has not reached widespread use among general practitioners (GPs), except in Australia [6]. In view of increased melanoma incidence rates, national screening programmes, and growing awareness of skin cancer in the public, there seems to be a need for clinical decision support systems (CDSSs) for GPs not familiar with dermoscopy.

For the past few decades, many CDSSs (also referred to as computer-aided diagnostic (CAD) systems) have been developed for melanoma detection. For a description of the major steps and an overview of different technologies, see, for example, [7]. The review paper of Rosado et al. [8] from 2003 concluded that the diagnostic accuracy of the CDSSs was statistically not superior to that of human diagnosis. Vestergaard and Menzies [9] reported the same in 2008, and Korotkov and Garcia [10] made a similar conclusion in 2012. Simultaneously high sensitivity and specificity scores have been reported [11], but the scores dropped when the same systems were tested in clinical-like settings with an independent test set of consecutively collected images. Dreiseitl et al. [12] concluded that the performance of their system in a clinical-like setting was significantly lower than the result obtained during training. Elbaum et al. [13] reported 100% sensitivity and 85% specificity on the training set, with a dramatic drop in specificity to 9% on an independent test set [14]. Also, Bauer et al. reported better results for the training set [15] than those achieved with an independent test set [16]. The good results of Hoffmann et al. [17] were reproduced in a small study with only 6 melanomas [18] but declined in another small study [19]. In two other small studies with independent test sets [20, 21], the performance reported by Blum et al. [22] declined. There are several possible explanations for the drop in performance. In studies where cross-validation has been used to validate the performance, the whole data set is used for feature selection or model selection, which gives overly optimistic results [23, 24]. Also, several studies exclude non-melanocytic lesions post hoc based on the pathology reports, which introduces bias.

Differentiating between melanocytic and nonmelanocytic lesions can be challenging even for experienced dermoscopy users [25], and it is recommended that a CDSS for melanoma detection can handle nonmelanocytic lesions as well [8, 11, 26]. A lesion is classified as suspicious if it resembles a melanoma, but lesions that resemble NMSC should also be classified as suspicious, especially when used by GPs [27], and not be classified together with nonsuspicious lesions.

In this report we present the performance of a CDSS for melanoma detection, *Nevus Doctor* (ND), when applied to both melanocytic and nonmelanocytic lesions. To our knowledge, this has not been previously performed for image-based CDSSs. Shimizu et al. [28] included BCC but excluded other NMSCs. Several studies have included nonmelanocytic lesions, but without reporting sensitivity to NMSC [14–17, 29–31] or with less than three NMSCs [20, 32]. Recently, non-image-based technologies for melanoma detection have included NMSC sensitivity in their reported findings [33, 34].

There is no apparent ranking of the CDSSs for melanoma detection. In practice, CDSSs can only be compared if tested on the same set of lesions, as done by Perrinaud et al. [32]. We therefore compare the performance of ND to that of a commercially available system, *Mole Expert* (ME), on the same set of lesions. This methodology potentially identifies the diagnostic difficulty of the data set, supplementing the information on the proportion of melanomas in situ, Breslow depth, clinical diagnoses, and so forth. It also allows for indirect comparison of CDSSs if, in the future, some study chose to compare another system to ME.

## 2. Materials and Methods

*2.1. Data.* From March to December 2013, patients were recruited at a private dermatology practice in Pulheim, Germany. Adult patients scheduled for excision of a pigmented skin lesion were eligible for inclusion. Furthermore, patients who were having nonpigmented skin lesions excised were eligible for inclusion if melanoma, BCC, or SCC was a potential differential diagnosis. Patients attending the German skin cancer screening programme [35] were also eligible for inclusion if they had skin lesions selected for excision. Informed written consent was obtained from all patients prior to inclusion. Skin lesions were excised because of concern about malignancy or when requested by the patient for other reasons. All skin lesions were photographed prior to excision with a digital camera (Canon G10, Canon Inc., Tokyo, Japan) with an attached dermoscope (DermLite FOTO, 3Gen LLC, California, USA) and with a videodermoscope (DermoGenius ultra, DermoScan GmbH, Regensburg, Germany). All excised lesions were examined by a dermatopathologist. In the case of a malignant diagnosis, a second dermatopathologist examined the excised lesion and a consensus diagnosis was set.

*2.2. Automatic Image Analysis.* ND takes a dermoscopic image from the Canon/DermLite device as input and classifies the lesion. ND is still in an experimental phase. In a previous study, ND performed as well as three independent dermatologists in terms of melanoma sensitivity and specificity [36]. In another study, ND performed as well as an independent dermatologist on an independent test set consisting of 21 melanomas and 188 benign lesions [37]. The data set in this study partly overlaps with the test set in Møllersen et al. [37], but not with the training set. ND has not been retrained, so the present data set is independent. ND outputs a probability of malignancy for each lesion image, and the sensitivity can be tuned with a parameter $\alpha$ (see [36–38] for details).

ME (MoleExpert micro Version 3.3.30.156) takes a dermoscopic image from the DermoGenius device as input. The output is a number between $-5.00$ and $5.00$, where high values indicate suspicion of melanoma, and the sensitivity can be tuned by adjusting a threshold $t$. ME is intended for use on melanocytic lesions only. ME was chosen as the comparison system due to availability and the fact that it has been tested and compared to other commercial systems in clinical settings on a small scale [29, 32]. To our knowledge, the study of Perrinaud et al. [32] is the only study that compares several systems on the same set of lesions, and it is therefore not possible to pick the most adequate reference system.

*2.3. Statistical Analysis.* All excised lesions for which a pathology report was available were included in the analysis. Cases were excluded if either of the CDSSs could not give an output or if images were missing. Presence of hairs and bubbles, lesion size, inadequate segmentation, and so forth were not used as exclusion criteria. The data was divided into four classes according to the histopathological diagnosis: melanoma, NMSC, benign melanocytic lesions, and benign nonmelanocytic lesions. The melanoma class includes lesions where malignancy could not be ruled out by the dermatopathologists (ICD-10 D48.5), and the NMSC class includes precancerous lesions, as done in other studies [33, 34]. Precancerous lesions should be classified as suspicious, since the patient should receive treatment or follow-up. The sensitivity and specificity scores of ND and ME were calculated by adjusting parameters $\alpha$ and $t$ and classifying the lesions accordingly. Sensitivity refers to the ratio of malignant lesions classified as suspicious to the total number of malignant lesions. Since there are two classes of malignant lesions, the terms melanoma sensitivity and NMSC sensitivity are used for clarification when needed.

The clinical diagnoses are taken from the dermatologist's referrals to pathology. The clinical malignant class includes lesions where malignancy is a differential diagnosis. The clinical benign class consists of lesions that were given a benign diagnosis by the dermatologist but were referred to pathology, which indicates that malignancy could not be ruled out by the dermatologist, even if this was not explicitly stated.

## 3. Results and Discussion

*3.1. Results.* There were 516 consultations (47% women) included in the study and a total of 877 excised lesions. The minimum age was 18, the maximum age was 93, and median age was 53 years. Table 1 shows the histopathological diagnoses. The ratio of benign lesions per melanoma was 16 : 1, which is within the range for dermatologists [39, 40]. In total, 5% of the lesions were melanomas. The median Breslow depth for the 23 invasive melanomas was 0.50 mm, and the maximum Breslow depth was 2.25 mm. Table 2 shows the diagnoses and the Breslow depths according to the 2009 American Joint Committee on Cancer (AJCC) staging [41]. About 70% of the melanomas and about 90% of the NMSCs were clinically diagnosed as malignant. One lesion lacked clinical diagnosis. Of the 875 lesions with histopathological diagnosis, four were excluded because ME did not give an output (one naevus, one seborrheic keratosis, one BCC, and one SCC) and one was excluded because the Canon/DermLite image was lost (melanoma in situ), which corresponds to less than 1% of the images. In comparison, 10% of the lesions were excluded due to lesions size or device malfunctions in the study by Malvehy et al. [33].

Figure 1(a) shows receiver operating characteristic (ROC) curves for ND and ME. The red and blue solid curves show sensitivity versus specificity for the whole data set where the malignant class includes both melanoma and NMSC, and a clear distinction between ND and ME can be seen. The pink and turquoise dotted curves show sensitivity versus specificity for melanocytic lesions only, and there is no significant difference between ND and ME. Figure 1(b) shows NMSC sensitivity as a function of melanoma sensitivity.

The boxplots in Figure 2 illustrate each CDSS's ability to discriminate between the different classes of lesions. For both

TABLE 1: Histopathological diagnoses for the 877 skin lesions.

| Histopathological diagnosis | Total: 877 |
| --- | --- |
| **Benign lesions** | **727** |
| Melanocytic lesions | 596 |
| Naevus | 574 |
| Naevoid lentigo | 9 |
| Blue naevus | 8 |
| Spitz naevus | 4 |
| Sutton naevus | 1 |
| Nonmelanocytic lesions | 118 |
| Seborrheic keratosis | 80 |
| Lentigo senilis | 9 |
| Neurofibroma | 5 |
| Dermatofibroma | 8 |
| Hemangioma | 4 |
| Verruca | 2 |
| Acanthoma | 3 |
| Papilloma, fibroma molle, sebaceous gland hyperplasia, scar, eczema, folliculitis, and discoid lupus erythematodes | 7 |
| Collision tumours | 13 |
| **Malignant and precancerous lesions** | **148** |
| Melanoma | 45 |
| Invasive melanoma | 25 |
| Melanoma in situ | 20 |
| Nonmelanoma skin cancers | 103 |
| Adnexal carcinoma | 1 |
| Actinic keratosis | 13 |
| Basal cell carcinoma | 71 |
| Squamous cell carcinoma | 7 |
| Bowen's disease | 11 |
| **No histopathological diagnosis** | **2** |

TABLE 2: Characteristics of the 45 melanomas.

| Diagnosis/Breslow | Number |
| --- | --- |
| In situ | 13 |
| Lentigo maligna | 7 |
| Cutaneous metastasis | 2 |
| ≤1.00 mm | 19 |
| 1.01–2.00 mm | 3 |
| 2.01–4.00 mm | 1 |
| >4.00 mm | 0 |

ND and ME, there is an overlap between melanomas in situ and benign melanocytic lesions, whereas invasive melanomas can be separated from benign melanocytic lesions. ND has high scores for NMSC, but also for benign nonmelanocytic lesions. ME has greater relative variety for all categories than ND has which lead to larger overlaps and thus more difficulty in distinguishing the different categories.

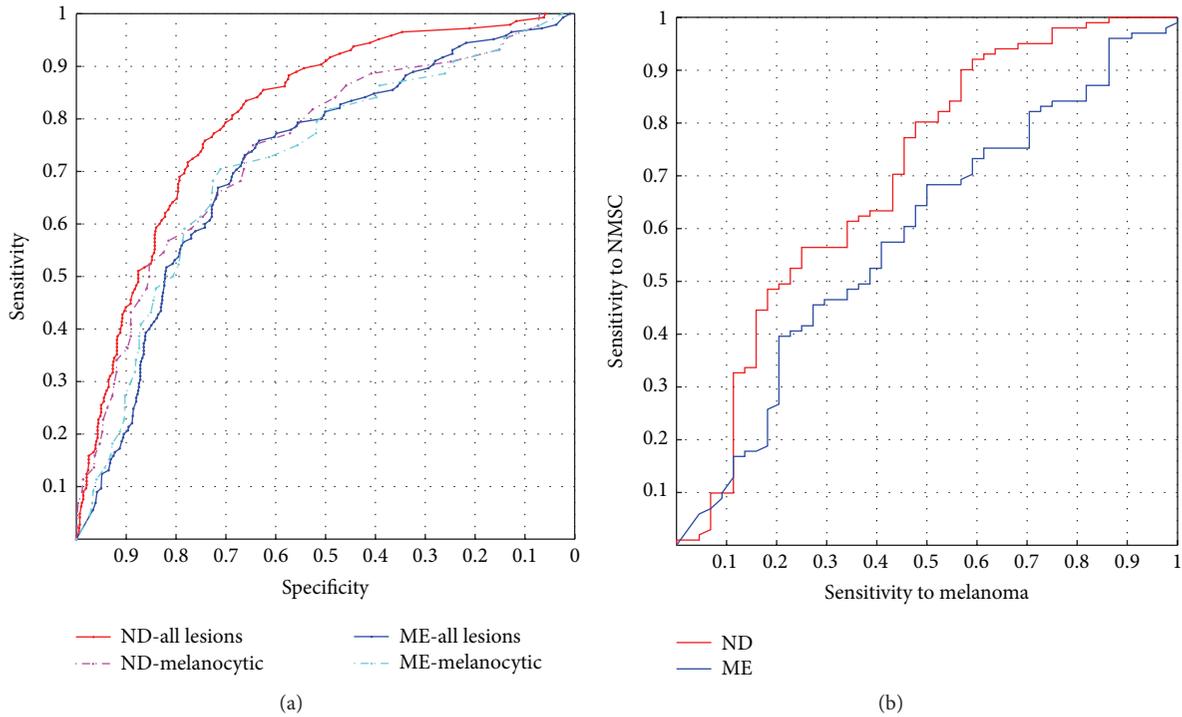

Figure 1: (a) Receiver operating characteristic curves for Nevus Doctor and Mole Expert. (b) Nonmelanoma skin cancer sensitivity as a function of melanoma sensitivity.

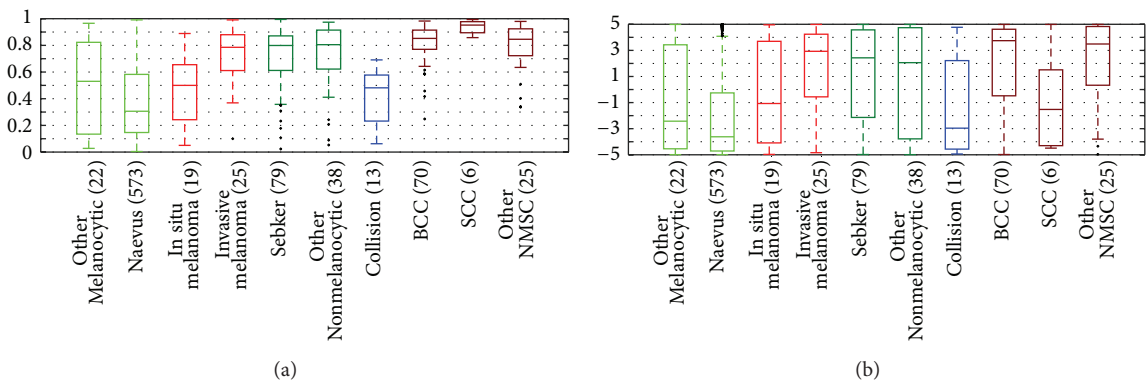

Figure 2: Boxplots for (a) Nevus Doctor and (b) Mole Expert. The horizontal lines inside the boxes are the medians; the boxes are defined by the 25th and 75th percentiles. The whiskers are situated at two standard deviations from the mean. The crosses indicate outlier observations.

For ND at 95% melanoma sensitivity, the overall sensitivity was 99%, the specificity was 12%, the positive predictive value (PPV) was 18%, and the negative predictive value (NPV) was 98%, where sensitivity, specificity, PPV, and NPV are defined in terms of malignant (melanoma and NMSC) and benign histopathological diagnosis. Figure 3 shows the two melanomas that were misclassified at 95% melanoma sensitivity for each CDSS. They were all in situ and were all clinically diagnosed as benign.

*3.2. Discussion.* The ROC curves for ND and ME in Figure 1(a) show that ND performed similar to ME when nonmelanocytic lesions were excluded, and by that we have shown that ND performed similarly to ME under the circumstances for which ME is intended. When non-melanocytic lesions were included, ND performed better than ME. Figure 1(b) shows that ND's NMSC sensitivity reached 100% at melanoma sensitivity of 86%, which means that, at reasonably high melanoma sensitivity, all NMSCs were classified as suspicious.

The majority of the benign lesions were excised due to suspicion of malignancy, and about 30% of the melanomas had a benign clinical diagnosis, so the overlap between benign melanocytic lesions and melanomas in situ, as seen in Figure 2, was expected. ND misclassifies seborrheic keratoses as suspicious, similar to other CDSSs [20, 31–33]. GPs excise more seborrheic keratoses per melanoma than dermatologists do [39, 40, 42], so there is a potential for saving pathology resources if a CDSS can classify them correctly. Development of image analysis features to discriminate seborrheic keratosis from melanoma has just begun [28].

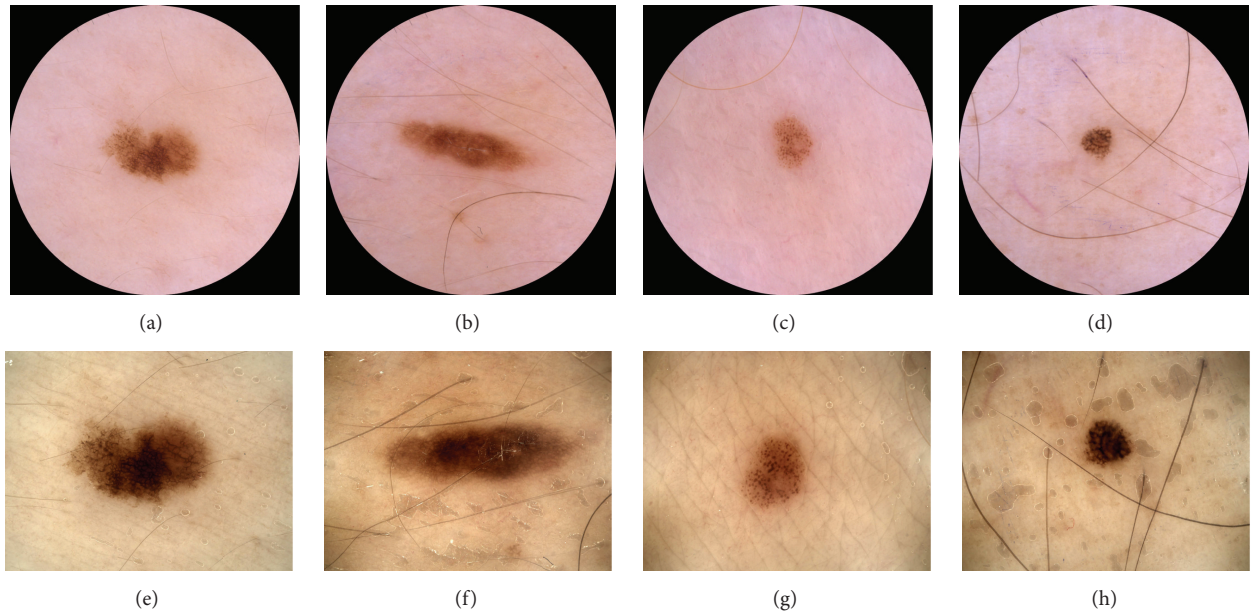

Figure 3: (a)–(d) Lesions photographed with Canon/DermLite. (e)–(h) Lesions photographed with DermoGenius. ((a)-(b) and (e)-(f)) The two melanomas misclassified by ND at 95% melanoma sensitivity. ((c)-(d) and (g)-(h)) The two melanomas misclassified by ME at 95% melanoma sensitivity.

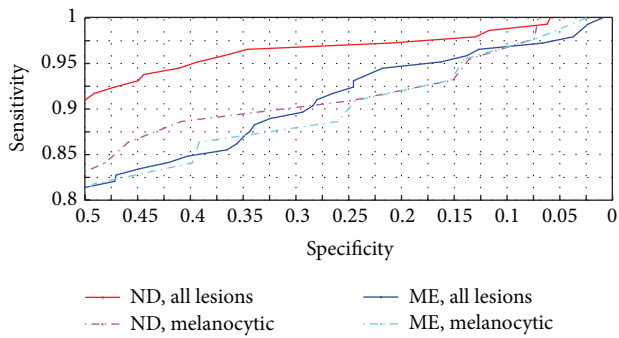

Figure 4: Receiver operating characteristic curves for Nevus Doctor and Mole Expert. Upper right part of Figure 1(a).

Only the segment of the ROC curve with high sensitivity to melanoma, as shown in Figure 4, is clinically relevant, and therefore the area under the curve (AUC) is not an adequate summary of a CDSS's performance. The area under the clinically relevant segment of the ROC curve would be a more adequate summary statistic; however, then the clinically relevant segment must first be defined.

Inadequate segmentation was not used as exclusion criterion, although it seems unlikely that a user would trust the outcome if the segmentation fails. But since there is no ground truth for segmentation, it is less suited as exclusion criterion. The decision to carry out excision was based on only one dermatologist's opinion, which is a drawback since interobserver agreement is only moderate for dermatologists [25]. Short-term follow-up of the patient or consensus diagnosis based on dermoscopic and clinical images can be used as the gold standard for nonexcised lesions, but it requires large resources. The data set used in this study is independent of all stages of ND's development, but the lesions in this set are possibly more similar to the lesions ND has been trained on than to the lesions ME has been trained on, and this is potentially an advantage for ND.

The present data set consisted of 44% melanomas in situ and median Breslow depth of 0.50 mm, which indicate high diagnostic difficulty. A proportion of 31% melanomas in situ was reported on a similar population in Germany [43]. The data set consists of 44 melanomas, which is more than most reported studies of CDSSs with independent test sets, with some exceptions [14, 33]. With 44 melanomas, 95% melanoma sensitivity means that only two melanomas are misclassified, and the results are therefore very sensitive to small changes in the data set. Confidence intervals for highly data sensitive results can give a false impression of generalisability and are therefore not reported.

ND and ME misclassified different melanomas, as shown in Figure 3, which is not unexpected since the two CDSSs have been developed independently of each other. It has been reported that the sensitivity and specificity for dermatologists improve when majority vote or consensus is used to diagnose skin lesions [5, 25]. We have investigated whether a combination of ND and ME will increase the sensitivity without decreasing the specificity. Figure 5 shows the ROC curve for a classifier which classifies a lesion as suspicious if either ND or ME classifies it as suspicious. For melanocytic lesions only, this classifier outperformed ND and ME for sensitivities above 90%. For melanoma and NMSC, the performance was about the same as for ND (not shown). An explanation of the results can be that the lesions were photographed with different cameras and dermoscopes, but it can also be because of the different feature algorithms or different statistical classifiers.

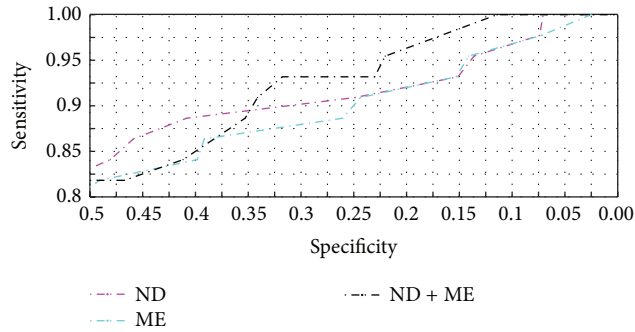

FIGURE 5: Receiver operating characteristic curves for Nevus Doctor, Mole Expert, and the combination classifier. Melanocytic lesions only.

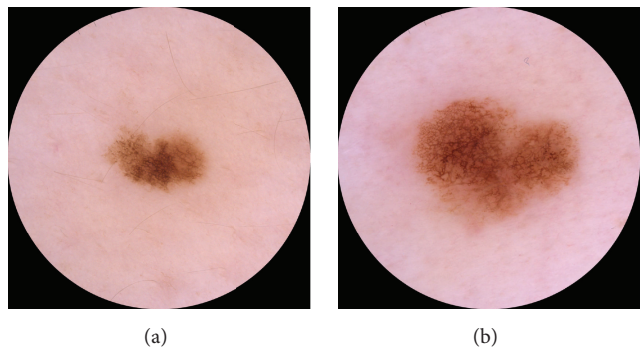

(a)                                        (b)

FIGURE 6: Lesion (a) was clinically diagnosed as naevus, whereas lesion (b) was clinically diagnosed as melanoma.

A CDSS is not needed for skin lesions that obviously are melanomas, and it can therefore be argued that these lesions should be excluded [14]. To evaluate ND's performance under these conditions, all melanomas that were clinically diagnosed as melanoma with no benign differential diagnosis were excluded. There were then 22 melanomas left. Tuning the parameters to 95% melanoma sensitivity (21 of 22 melanomas detected) gave a specificity of 6% for ND and 13% for ME, similar to the 9% specificity reported by Monheit et al. [14]. One should, however, be very cautious when drawing conclusions. To demonstrate the dependence on the data set, one of the melanomas clinically diagnosed as benign was replaced by a melanoma clinically diagnosed as malignant, shown in Figure 6. The specificity then increased to 12% for ND and decreased to 7% for ME, which is outside the range of the respective confidence intervals.

Compared to other studies with independent test sets and a minimum of 10 melanomas [12, 14, 16, 33, 34], the sensitivity and specificity scores of ND and ME are not superior. Inclusion of nonexcised lesions [12, 34] can have a positive effect on the observed specificity, since these lesions do not resemble malignant lesions. A higher proportion of melanomas in situ decreases the observed performance of the system as shown in Figure 2 and also illustrated by Malvehy et al. [33]. Whether the superior performance of Piccolo et al. [16] is due to few melanomas in situ is unknown, as this was not reported. The two studies with more than 100 melanomas in an independent test set reported similar sensitivities (98% and 97%) but different specificities (9% and 34%), but it is not possible to conclude that one system is better than the other, since the exclusion criteria for the two studies are very different. A publicly available data set has been called for [8], but the wide variety of technologies is a challenge.

## 4. Conclusions

We have shown that ND was able to detect NMSC without sacrificing melanoma sensitivity but misclassified benign nonmelanocytic lesions. Although there are promising results for other technologies, dermoscopy is still the only widely used tool for skin lesion diagnosis. Nonmelanocytic lesions are an important aspect in melanoma detection, and more research is needed, especially on features for differentiating between melanomas and seborrheic keratoses.

Different inclusion and exclusion criteria, moderate-sized data sets, and variety in the diagnostic difficulty make the reported sensitivity and specificity scores inadequate for comparing different CDSSs. The demonstration of the results' dependence on the data set emphasises the need for direct comparison on the same set of lesions. Which system is the most adequate for comparison will remain unknown until more studies are reported.

According to a study by Dreiseitl and Binder [44], physicians are willing to follow the recommendation of a CDSS, especially if they are not confident in their own diagnosis, and Frühauf et al. [20] reported that patients accept the use of a CDSS. Hence, there is a potential for CDSSs in melanoma detection if the systems can give reliable recommendations for all kinds of lesions.

## Conflict of Interests

The authors declare that there is no conflict of interests regarding the publication of this paper.

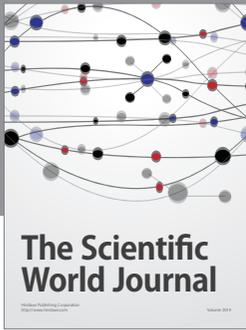

**The Scientific World Journal**

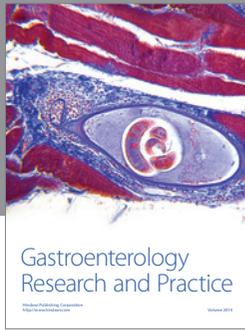

Gastroenterology Research and Practice

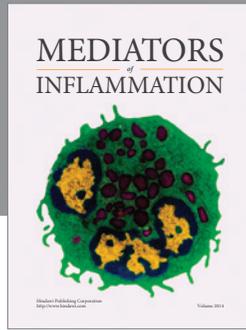

MEDIATORS of INFLAMMATION

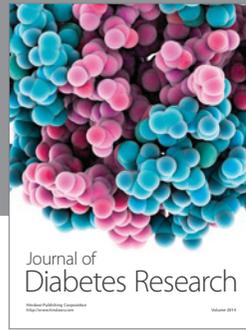

Journal of Diabetes Research

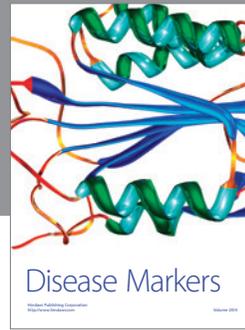

Disease Markers

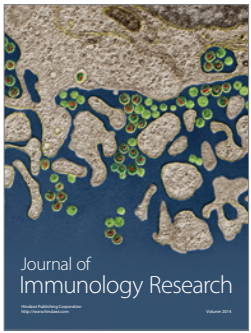

Journal of Immunology Research

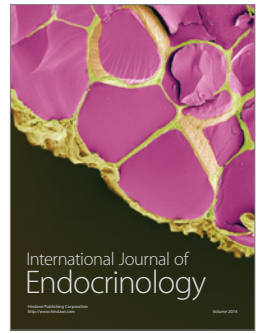

International Journal of Endocrinology

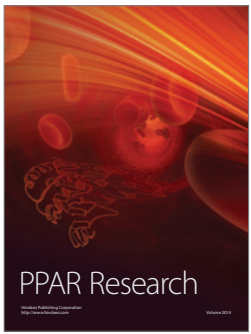

PPAR Research

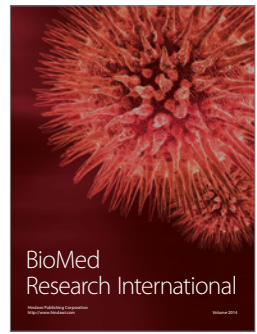

BioMed Research International

Hindawi

Submit your manuscripts at
http://www.hindawi.com

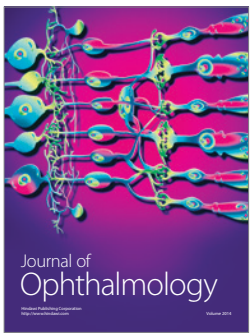

Journal of Ophthalmology

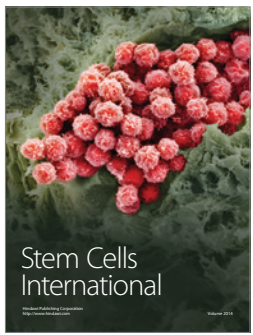

Stem Cells International

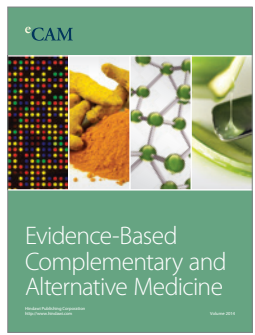

CAM

Evidence-Based Complementary and Alternative Medicine

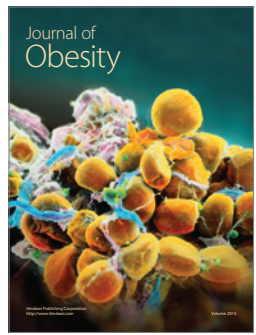

Journal of Obesity

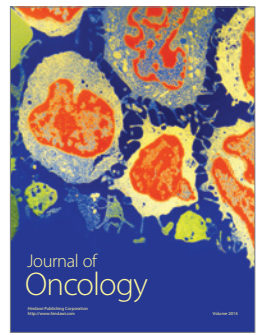

Journal of Oncology

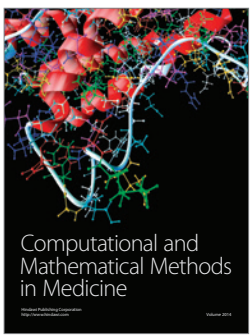

Computational and Mathematical Methods in Medicine

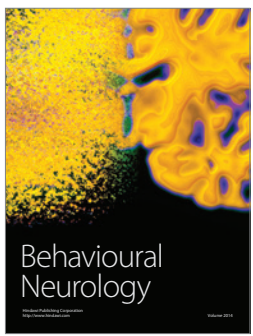

Behavioural Neurology

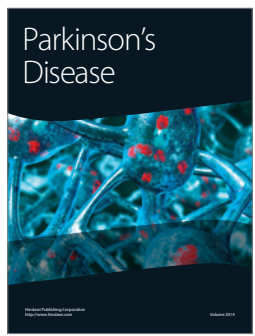

Parkinson's Disease

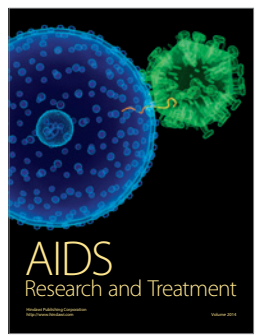

AIDS Research and Treatment

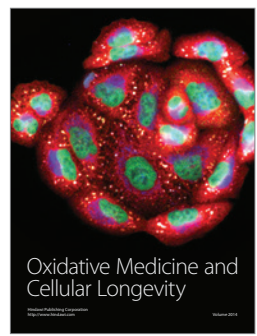

Oxidative Medicine and Cellular Longevity